\documentclass[a4paper]{article}

\usepackage[english]{babel}
\usepackage[utf8]{inputenc}
\usepackage{graphicx}%
\usepackage{multirow}%
\usepackage{amsmath,amssymb,amsfonts}%
\usepackage{amsthm}%
\usepackage{mathrsfs}%
\usepackage[title]{appendix}%
\usepackage{xcolor}%
\usepackage{textcomp}%
\usepackage{manyfoot}%
\usepackage{booktabs}%
\usepackage{algorithm}%
\usepackage{algorithmicx}%
\usepackage{algpseudocode}%
\usepackage{listings}%
\usepackage{xspace}%
\usepackage{units}
\usepackage[normalem]{ulem}
\usepackage{url}
\usepackage{hyperref}

\def\p{\mathbf{p}}
\def\x{\mathbf{x}}

\newcommand{\eVnospace}{\text{e\kern-0.15ex V}\xspace}
\newcommand{\eV}{\text{ e\kern-0.15ex V}\xspace}

\newcommand{\GeV}{\text{ G\eVnospace}\xspace}
\newcommand{\TeV}{\text{ T\eVnospace}}

\newcommand{\vtwo}{\ensuremath{{v_{2}}}\xspace}
\newcommand{\meanvtwo}{\ensuremath{\left\langle v_{2} \right\rangle}\xspace} 

\newcommand{\pythia}{\textsc{Pythia}\xspace}

\newcommand{\appropto}{\mathrel{\vcenter{
  \offinterlineskip\halign{\hfil$##$\cr
    \propto\cr\noalign{\kern2pt}\sim\cr\noalign{\kern-2pt}}}}}

\newcommand{\ie}{\textit{i.e.}\xspace}
\newcommand{\eg}{\textit{e.g.}\xspace}
\newcommand{\cf}{\textit{cf.}\xspace}

\begin{document}

\begin{center}
  \textbf{\LARGE Going against the flow}\\
  \large{Revealing the QCD degrees of freedom in hadronic collisions}
\end{center}

\begin{center}
  Christian Bierlich\textsuperscript{1$\spadesuit$},
  Peter Christiansen\textsuperscript{1$\dagger$},
  Gösta Gustafson\textsuperscript{1$\star$},
  Leif Lönnblad\textsuperscript{1$\maltese$},
  Robin Törnkvist\textsuperscript{2$\ddagger$},
  Korinna Zapp\textsuperscript{1$\perp$},
\end{center}

\begin{center}
  \textbf{\textsuperscript{1}}
  Department of Physics, Lund University, Box 118, SE-221 00 Lund, Sweden \\
  \textbf{\textsuperscript{2}}
  Instituto Galego de Física de Altas Enerxía, Universidade de Santiago de Compostela, Santiago de Compostela, ES-15782, Spain
  ${}^\spadesuit$\textsf{\small christian.bierlich@fysik.lu.se},
  ${}^\dagger$\textsf{\small peter.christiansen@fysik.lu.se},
  ${}^\star$\textsf{\small gosta.gustafson@fysik.lu.se},
  ${}^\maltese$\textsf{\small leif.lonnblad@fysik.lu.se},
  ${}^\ddagger$\textsf{\small robin.tornkvist@usc.es},
  ${}^\perp$\textsf{\small korinna.zapp@fysik.lu.se}.
\end{center}

\section*{Abstract}
\addcontentsline{toc}{section}{\protect\numberline{}Abstract}
In collisions between heavy nuclei, such as those at the Large Hadron Collider (LHC) at CERN, hydrodynamic models have successfully related measured azimuthal momentum anisotropies to the transverse shape of the collision region. For an elliptically shaped interaction area, the hydrodynamic pressure gradient is greater along the minor axis, resulting in increased particle momentum in that direction --- a phenomenon known as \emph{positive} elliptic flow.

In this paper, we demonstrate that in smaller systems, such as proton--proton and peripheral ion--ion collisions, microscopic models for final state interactions, can produce anisotropies where the elliptic flow is \emph{negative} --- that is, the momentum is largest along the major axis, contrary to hydrodynamic predictions.

We present results from two distinct microscopic models: one based on repulsion between string-like fields and another based on effective kinetic theory. Negative elliptic flow is a solid prediction of the string interaction model while in the model based on kinetic theory it is linked to a finite interaction range. Consequently, an experimental determination of the sign of elliptic flow, will provide novel insights into the degrees of freedom governing strong nuclear interactions in high-energy collisions and the way in which they interact.

\begin{center}
\textbf{Keywords:} Particle Physics, Heavy Ion Physics, Quantum Chromodynamics
\end{center}
\newpage
\section*{Introduction}
\addcontentsline{toc}{section}{\protect\numberline{}Introduction}
Proton--proton (pp) collisions are among the most powerful tools for probing the microscopic nature of our Universe, best illustrated by the discovery of the Higgs boson at the Large Hadron Collider (LHC) in 2012~\cite{ATLAS:2012yve,CMS:2012qbp}. Each pp collision typically involves multiple subcollisions between quarks and gluons. In searches for new particles, such as the Higgs boson or physics beyond the Standard Model, the hardest interaction --- where the largest momentum transfer occurs --- is of primary interest. The hardest scattering can be understood from first principles, given that the theory of the strong interaction, Quantum Chromodynamics (QCD), is asymptotically free at high momentum transfers~\cite{Gross:1973id,Politzer:1973fx}.

The softer interactions involving low momentum transfers between the remaining quarks and gluons are less well understood, as perturbation theory loses its applicability. These softer processes constitute the so-called ``underlying event'', which is typically considered a background in searches for new particles. However, the past decade of LHC data has revealed unexpected and intriguing physics, anisotropic flow~\cite{CMS:2010ifv,ALICE:2012eyl,PHENIX:2018lia} and strangeness enhancement~\cite{ALICE:2016fzo}, within this underlying event.

In heavy-ion collision, anisotropic collective flow~\cite{Ollitrault:1992bk} and enhanced production of the heavier strange  quarks~\cite{Cabibbo:1975ig,Rafelski:1982pu} are signatures widely interpreted to indicate the formation of a Quark--Gluon Plasma (QGP) --- a state of matter that existed in the early Universe shortly after the Big Bang. This interpretation is founded in non-perturbative lattice QCD calculations \cite{Bazavov:2011nk}. QGP formation is facilitated by strong re-scattering effects in the final state, \ie, after the first generation of particles has been formed. The aforementioned discoveries of QGP signatures in small systems, has called this paradigm into question. Here, confinement based phenomenological models, without sizable final state re-scattering effects, have traditionally been very successful. The field is thus faced with fundamental questions about the dynamics of small collision systems.

Recently, measurements of plasma-like behaviour in small, high-energy collision systems, has been complemented by measurements of emergent hydrodynamic behaviour in cold atom systems \cite{Brandstetter:2023jsy}. As this is now suggested as an analogous system to the much smaller proton collisions, it is more than timely to ask whether hydrodynamic behaviour is really universal, or if limiting cases occur when the system is small enough that microscopic degrees of freedom must be taken into account. 

\begin{figure}[h]
\centering
\includegraphics[width=\textwidth]{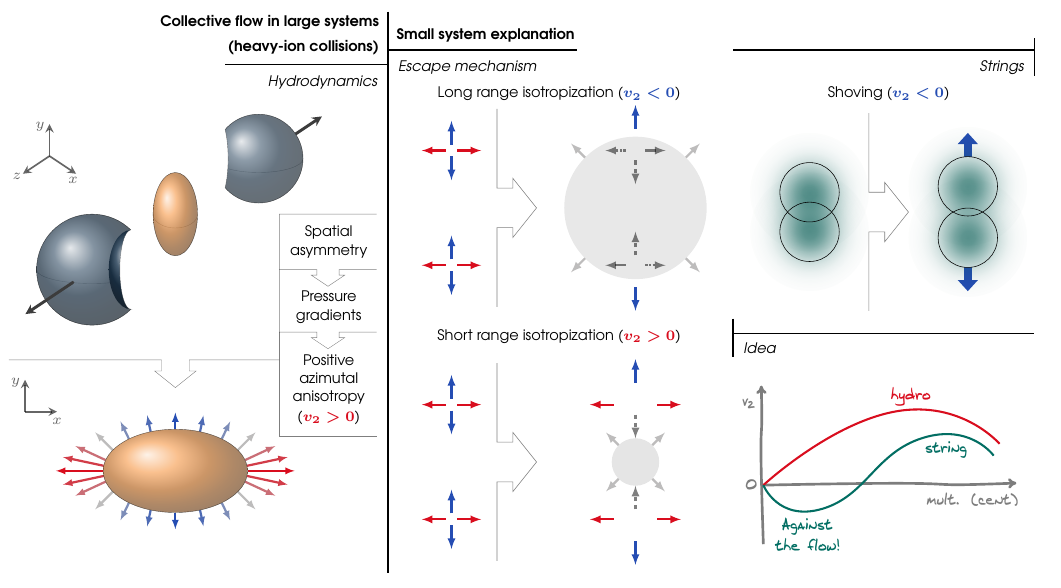}
\caption{Illustration of how the sign of the elliptic flow, \vtwo, varies for
    different systems and the models considered here. Particles indicated with red arrows, 
    give a positive contribution to \vtwo, while
    particles indicated with blue arrows contribute negatively. Final-state
    particles with no (average) contribution to \vtwo (isotropization)
    are grey. \textit{Left}: For large systems, the elliptic flow is generated
    by hydrodynamic pressure gradients, stronger along the minor
    axis. \textit{Center top}: For kinetic theory, when system size is small compared to
    the interaction range (indicated by grey circles), the escape mechanism gives a negative
    \vtwo. \textit{Center bottom:} In the opposite case it gives positive
    \vtwo. \textit{Right}: Two parallel strings repel each other, giving
    negative \vtwo.}\label{fig1}
\end{figure}

In this paper, we employ two distinct microscopic models to study anisotropic flow in such collisions. In the large-system limit, both models converge to hydrodynamic behaviour \cite{Kurkela:2018qeb,Bierlich:2020naj}. We will first provide qualitative arguments for the main results of the paper along the lines of Fig.~\ref{fig1}, depicting the main differences between the models.

\section*{Background and main thesis}
\addcontentsline{toc}{section}{\protect\numberline{}Background and main thesis}
In non-central Pb--Pb collisions, the shape of the region where the nuclei overlap is elliptic in the plane transverse to the beam direction (Fig.~\ref{fig1}, left). For such collisions, the final state evolution is well described by nearly ideal \textit{relativistic hydrodynamics} \cite{Heinz:2013th}. Pressure gradients drive the system’s expansion, which occurs more rapidly along the minor axis of the ellipse, leading to a \emph{positive} elliptic flow response. The elliptic flow response is quantified in the coefficient \vtwo, which will be introduced later.

In smaller collision systems, \textit{kinetic theory} provides a different mechanism for generating anisotropic flow (Fig.~\ref{fig1}, center). Kinetic theory predicts macroscopic properties and transport phenomena based on microscopic interactions between particles, even when the system is out of equilibrium and the mean-free path is not negligible compared to the system size. 
At high temperatures, the QGP consists of quarks and gluons as well-defined quasi-particles. The AMY (Arnold, Moore, and Yaffe) theory \cite{Arnold:2002zm} formulates Boltzmann equations for (anti)quarks and gluons, describing the phase space densities $f_s(\x,\p,t)$ (for particles of species $s$ with momentum $\p$ at point $\x$) due to free streaming and interactions. We focus on elastic $2 \leftrightarrow 2$ scattering, which is the dominant process for generating flow-like correlations. Quasi-collinear $1\leftrightarrow 2$ splitting or merging processes, also present in QCD, do not significantly alter particle directions and contribute only minor corrections.

We use a parton cascade approach (\cf~appendix \ref{app2}) to solve the Boltzmann equation, where particles interact over a finite distance defined by the scattering cross section. Kinetic theory generically produces a positive flow-like response when the system size is large compared to the interaction range \cite{Kurkela:2018ygx}. In such cases, particles moving towards the center along the vertical axis in Fig.~\ref{fig1} (center bottom) interact and scatter, leading to an excess of particles moving horizontally and thus a positive elliptic flow \cite{Kurkela:2018ygx}. This is known as the \textit{escape mechanism}, and is a prime candidate for explaining positive \vtwo in collision systems out of equilibrium, as well as in cold atom systems. For systems not large compared to the interaction range (Fig.~\ref{fig1} center top), most particles scatter except those moving away from the center along the vertical axis. This can result in a \textit{negative} elliptic flow.

In the \textit{Lund string hadronization model} \cite{Andersson:1979ij,Andersson:1983jt,Andersson:1983ia} it is assumed that the confining interaction between quarks and anti-quarks can be described by flux
tubes similar to vortex lines in a superconductor, as originally suggested in Refs.~\cite{Nielsen:1973cs,Nambu:1974zg}. When the string width is not important, \eg, in systems with just a single, straight string, the dynamics corresponds to a massless relativistic string (a
Nambu--Got\={o} string)~\cite{Nambu:1970,Goto:1971ce}. The string breaks up by production of $q\bar{q}$ pairs, in analogy to the production of e$^+$e$^-$ pairs in 1+1-dimensional QED~\cite{Schwinger:1962tp}. The string model is a core part of the free open source \pythia Monte Carlo Event Generator program~\cite{Bierlich:2022pfr} (\cf~appendix \ref{app1}). 

When several proton constituents interact, more strings are produced, and the width of the string flux tubes can no longer be neglected. Parallel strings then repel each other analogous to the interaction
between vortex lines in a superconductor~\cite{Bierlich:2020naj}. In Fig.~\ref{fig1} (right top) we illustrate an idealised low-multiplicity event with two strings within an elliptical interaction region. The repulsion between these strings results in expansion along the vertical axis, leading to a \textit{negative} elliptic flow. With an increased number of strings, the system resembles the nucleus case more closely, resulting in a positive elliptic flow.

In Fig.~\ref{fig1} (right bottom), we illustrate how these two models predict a negative elliptic flow in small collision systems. The size of the collision system is represented by multiplicity of particles produced in the event in arbitrary units. Both models can, in principle, produce a negative elliptic flow response if the initial state is suitably engineered, whereas a hydrodynamic response will never do so.

\section*{Measuring the sign of $v_2$}
\addcontentsline{toc}{section}{\protect\numberline{}Measuring the sign of $v_2$}
We will briefly review the methods used to measure \vtwo, and thus how the sign of \vtwo is determined. 

In heavy ion collisions the distribution in azimuthal angle ($\varphi$) of final-state particles is approximately proportional to 
\begin{equation} 
\label{eq:dndphi}
\frac{\mathrm{d}N}{\mathrm{d}\varphi} \appropto 1 + 2 \vtwo \cos[2 (\varphi -\Psi)],
\end{equation}
where $\Psi$ represents the impact-parameter angle, \ie, the line connecting the centres of the colliding nuclei in the transverse plane (horizontal axis in Fig~\ref{fig1}). 
For model calculations where $\Psi$ is known, \vtwo is determined by averaging over events:
\begin{equation}
  \label{eq:vtwo}
  \langle \vtwo \rangle = \langle \cos[2 (\varphi - \Psi)] \rangle.
\end{equation}
Experimentally, one can measure the magnitude of \vtwo without measuring $\Psi$. Instead it is determined from correlations between two (or more) particles, often referred to as $v^2_2\{2\}$:
\begin{equation}
\label{eq:vtwotwo}
\langle \cos[2  (\varphi_1 - \varphi_2)] \rangle = \langle v^2_2\{2\} \rangle \ (\neq \langle v_2 \rangle^2). 
\end{equation}
Estimates using Eq. (\ref{eq:vtwotwo}) have the best precision, as one avoids experimental uncertainties on the hard-to-measure $\Psi$, and non-flow correlations, \eg~from jets and resonance deceays, can be suppressed by using four or more particles. In collisions of nuclei, direct measurements of $\Psi$ from spectator nucleus remnants are rarely pursued. 
Instead the sign of $v_2$ has been determined by 3-particle correlations using the directed flow, $v_1$, of pions in the  forward region. With this method the sign of $v_2$ in high energy heavy ion collisions was determined to be positive in the NA49 experiment at the SPS \cite{NA49:2003njx} and the STAR collaboration at RHIC \cite{STAR:2003xyj}. In pp collisions this method is difficult because the directed flow is very small. Therefore, the sign of \vtwo in proton collisions remains experimentally unknown. In measurements of cold atom systems \cite{Brandstetter:2023jsy}, $\Psi$ can be measured directly, and there are no ambiguities regarding the sign of \vtwo.

\section*{Results}
\addcontentsline{toc}{section}{\protect\numberline{}Results}
\subsection*{Negative \vtwo in two- and four-hotspot systems}
\begin{figure}[h]
\centering
\includegraphics[width=\textwidth]{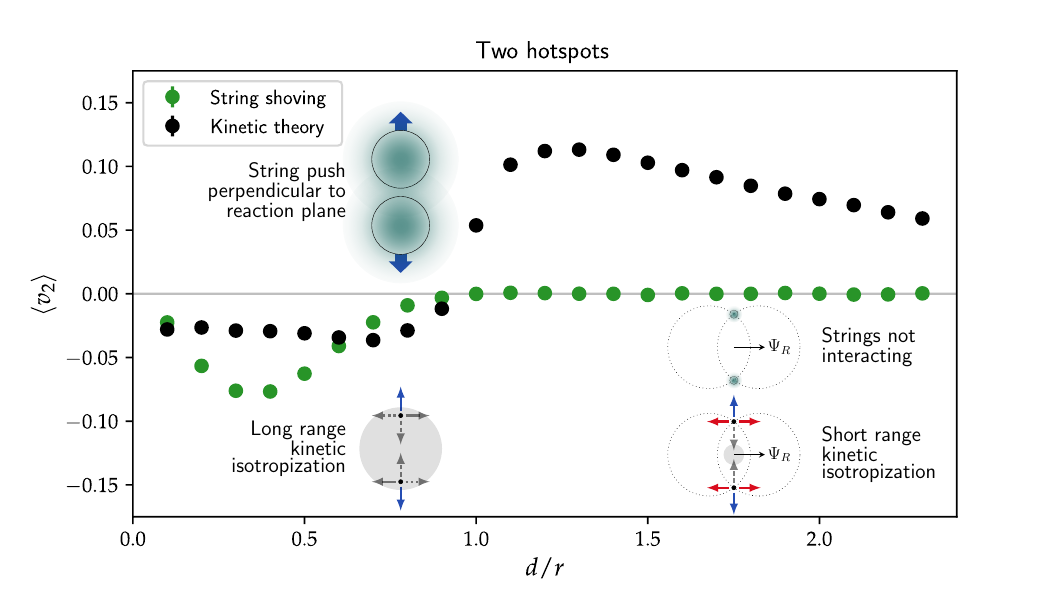}
\caption{Average elliptic flow dependence on distance ($d$) divided by interaction length ($r$), with a two-hotspot initial state. The two hotspots are positioned as indicated in the illustrations. The mechanism behind the sign of the flow in each region, $d/r < 1$ and $d/r > 1$, is indicated.} \label{fig2}
\end{figure}

\addcontentsline{toc}{subsection}{\protect\numberline{}Negative $v_2$ in two- and four-hotspot systems}
To demonstrate how the sign of the elliptic flow depends on system size, we first discuss a simple toy model with two hotspots. In kinetic theory, a ``hotspot'' refers to a source of particles. We restrict ourselves to consider only gluons. Each event has a total of 10 gluons, distributed randomly among the hotspots, with intrinsically isotropic momentum distribution. In the string model, a hotspot represents the center of a single, straight quark--antiquark string, with a total energy of 30 GeV. All energy is put in the quark's momentum in the $z$-direction, apart from a random kick of order 1 GeV in the $x$ and $y$ directions, to prevent numerical instabilities in the calculation. In both cases, hotspots are separated by a distance $d$ in the vertical direction, and Fig.~\ref{fig2} shows how \meanvtwo depends on the separation between the hotspots.

The distance $d$ is normalized to the interaction length, $r$, of the two models. In the string model, $r$ corresponds to twice the string radius, $\sqrt{\langle R^2\rangle} = \unit[1]{fm}$. In the kinetic theory calculation, $r$  is the characteristic size obtained from the fixed elastic scattering cross section $\sigma$, through $r=\sqrt{\sigma/\pi}$ with $\sigma = 4\pi$ fm$^2$ (which is unrealistically large for QCD, but is chosen here for illustrative purposes)

The string model result can be directly understood through the qualitative considerations shown in Fig.~\ref{fig1}. When the separation between strings is large enough that they do not overlap, there is no interaction or ``shoving'', resulting in \meanvtwo = 0. However, as soon as the strings begin to interact, the elliptic flow becomes negative. We note that for this geometry, the string model can never produce a positive $v_2$. It is also important to note that in the limit of no separation, the average elliptic flow will once again approach zero.

For kinetic theory, the result in Fig.~\ref{fig2} provides a quantitative estimate of the escape mechanism's effect, confirming the trends discussed earlier in connection with Fig.~\ref{fig1}. Crucially, \vtwo will only turn negative when the number of scatterings per particle exceeds unity. However, in this scenario, the mean free path is shorter than the interaction time, meaning that kinetic theory is formally not applicable. Nonetheless, it is conceivable that the actual microscopic theory would behave similarly. We point out that a finite interaction range is crucial for the appearance of negative $\vtwo$. When the kinetic theory is solved by direct (analytical or numerical) integration of the Boltzmann equation the phase space densities of the colliding particles are evaluated at the same space-time point, which corresponds to the limit of perfectly local interactions with vanishing interaction range. This corresponds to the short range kinetic isotropization scenario in Figs.~\ref{fig1} and \ref{fig2}, which will always give rise to a positive \vtwo. In kinetic theory the sign of \vtwo in small systems could thus yield qualitative insights about the nature of the interaction although the theory is not necessarily formally applicable.
  
\begin{figure}[h]
  \centering
  \includegraphics[width=\textwidth]{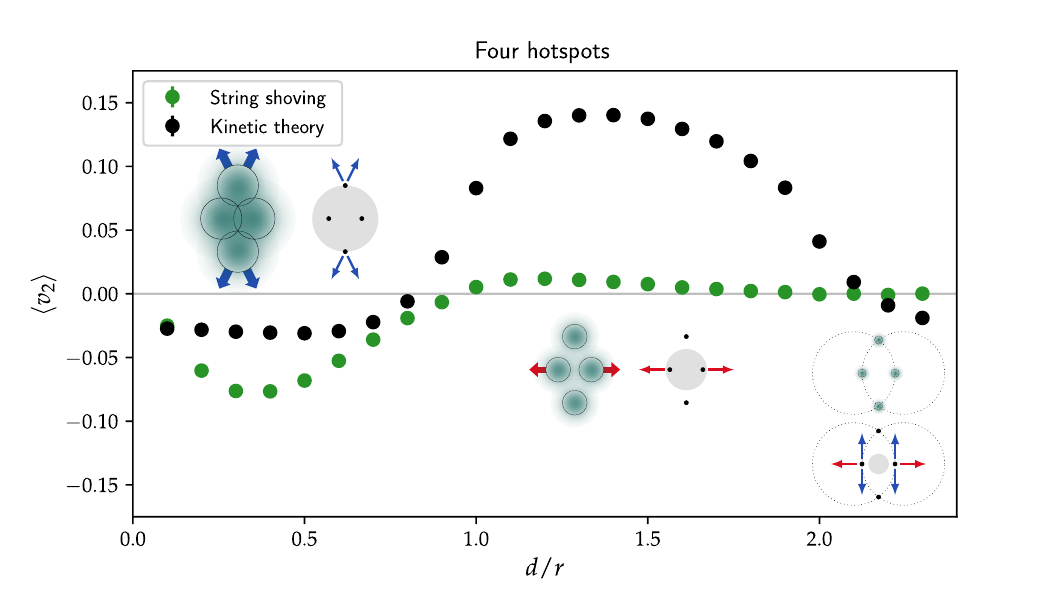}
  \caption{Average elliptic flow dependence on distance ($d$) divided by interaction length ($r$), with a four-hotspot initial state. Here $d$ denotes the vertical separation between the two central hotspots, the horizontal hotspots are separated by $d/2$. The figures in each region illustrate the dominant mechanism behind the sign of the flow in that region.} \label{fig3}
\end{figure}

A four-hotspot initial state provides a more realistic representation of a pp collision. In Fig.~\ref{fig3}, we consider a geometry constructed from the intersection points of two overlapping circles, which represent the projectile and target in a pp collision. The characteristic size ($d$) of the system is defined as the vertical separation between the two central hotspots. The dependence on $d$ is now more complex, but the models again predict negative \vtwo of comparable magnitude for small system sizes.

In the case of strings, the system behavior divides into three regions. When the system size is small (closely mimicking a pp collision, where the string size is comparable to the system size), all strings can interact, leading to a negative \vtwo. For larger values of $d$, the interaction with the two strings on the vertical axis can be neglected and the repulsion between the strings on the horizontal axis results in a positive \vtwo. For very large separations, \vtwo again tends to zero, as expected.

In kinetic theory, the escape mechanism for a small system isotropizes most particles, except those moving close to the vertical axis leading to negative \vtwo. For intermediate system sizes, the escape mechanism primarily affects the two inner hotspots, resulting in a positive \vtwo. For larger systems, mainly particles originating from the inner hotspots and moving toward each other randomize, leading to a return to a negative \vtwo.

\subsection*{Towards realistic collision systems}
\addcontentsline{toc}{subsection}{\protect\numberline{}Towards realistic collision systems}
While the two- and four-hotspot initial states serve well to illustrate the physical principles and predictions of the two microscopic models, they provide little certainty that these phenomena will also be present in realistic collision systems. In the following we present results for more realistic initial states, using state-of-the-art calculations with a saturation based calculation for the \textsc{Alpaca} kinetic theory calculation, and an initial state based on the \pythia model in the string calculation.
\begin{figure}[h]
\centering
\includegraphics[width=0.49\textwidth]{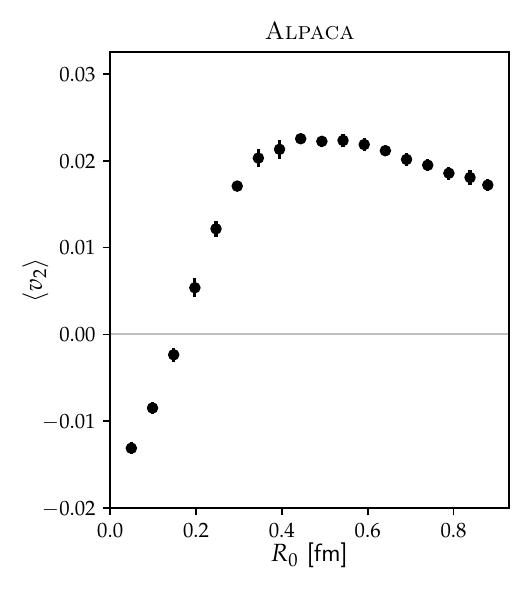}
\includegraphics[width=0.49\textwidth]{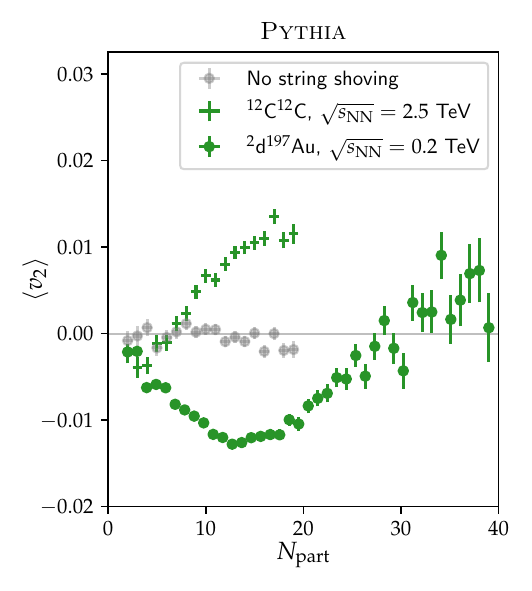}
\caption{Results for kinetic theory (left) and string interaction (right) \vtwo for realistic initial states. In the former case using a saturation based calculation showing \vtwo vs. $R_0$, system size. In the latter case for two select ion--ion initial states, which both produce a change of sign in \vtwo when shown vs. $N_\text{part}$, a proxy observable for system size.}\label{fig4}
\end{figure}

The main result of this section is that a negative $v_2$ can be obtained for realistic initial conditions. Moreover, a change of sign can be observed within a single collision system. That may be be easier to measure due to rapid changes in fluctuations in $\langle v_2^2 \rangle$, even if the sign cannot be measured directly.

A more realistic kinetic theory scenario for small systems is provided by the Monte Carlo event generator \textsc{Alpaca}~\cite{Tornkvist:2023kan} (under development), which implements the AMY kinetic theory. The initial conditions are chosen as a purely gluonic system based on gluon saturation~\cite{Lappi:2011ju, Kurkela:2015qoa, Kurkela:2021ctp}. Specifically, we use the values derived in Ref.~\cite{Kurkela:2021ctp} (with minor changes, \cf~appendix \ref{app2}), which correspond to pp collisions at $\sqrt{s}=5.02~\TeV$. When changing the system size $R_0$, the occupancy (which sets the energy density) is adjusted to keep the number of initial gluons fixed at 16 to avoid numerical instabilities.

Figure~\ref{fig4} (left) shows the resulting \vtwo as a function of $R_0$. Equation (\ref{eq:vtwo}) is used to obtain \vtwo, with the event plane fixed as the initial symmetry plane (see Eq.~\eqref{eq:CGClike}), \ie, $\Psi = 0$. A clear negative regime of \vtwo is observed for small system sizes, with a change of sign at $R_0 \approx 0.2$ fm. There are two main uncertainties regarding the validity of the AMY treatment in the regime of negative \vtwo. Firstly, the interaction time is larger than the mean free path. Secondly, AMY requires a clear separation of scales between the screening mass and the typical momentum of the system, which becomes questionable for low $R_0$. So while the results are mathematically correct, more work is required to ensure that AMY can be applied to describe the physics of these small systems.

For strings, the shoving model, as implemented in the \pythia event generator, is used. The event generator is capable of simulating collisions at high energies of almost all projectile and target combinations (\cf~appendix \ref{app1}). Collisions of composite particles such as protons and neutrons, are based on a model of multiple partons interactions \cite{Sjostrand:1987su}, where several constituents can collide. Collisions of nuclei are handled by the Angantyr \cite{Bierlich:2018xfw} model, see appendix \ref{app1}. Crucially, the Angantyr model does not introduce any QGP in such collisions, but attempts to generate all collective effects by string--string interactions.

Figure~\ref{fig4} (right) shows the resulting \vtwo as function of number of nucleons participating in the collision ($N_\text{part}$) for two different collision systems; $^{12}$C$^{12}$C and $^2$d$^{197}$Au. We calculate $\Psi$ in Eq. (\ref{eq:vtwo}) from the positions of strings spanning the plane transverse to the beam axis at the collision point of the nuclei (\ie, at mid-rapidity), \cf~appendix \ref{app1}. This allows for a fine-grained event-by-event estimate of the spatial energy density.
The result shows, for both systems, a clear transition from a negative to a positive \vtwo regime for increasing system size. We mention that this transition is not seen in pp, hence it is not shown here.

\section*{Conclusions}
\addcontentsline{toc}{section}{\protect\numberline{}Conclusions}
In this paper we have investigated elliptic flow in small collision systems using two distinct microscopic models, each capturing key aspects of the collective dynamics. We show that, unlike in large systems where hydrodynamic behavior dominates, the elliptic flow in small systems can have negative sign. Furthermore, we show that the transition between negative and positive \vtwo might be observable for specific, realistic initial states.

In the string interaction model, negative \vtwo in small systems is a very solid prediction. This to the extent that an experimentally measured purely positive \vtwo would rule out the model. For kinetic theory the situation is less clear-cut. Within its range of applicability, where the interaction time is small compared to the mean free path, it will always predict a positive \vtwo. When the interaction time is not small compared to the mean free path, the sign of \vtwo depends on the method by which the theory is solved, in particular on the presence or absence of a finite interaction range. Therefore, this regime --- although formally outside the region of applicability of kinetic theory --- could inform us about the nature of the interaction and how to best go beyond Boltzmann transport. Complementary measurements on cold atom systems, where size and interaction range can be varied and $\Psi$ measured directly, could further inform the puzzle about the applicability of kinetic theory in the small interaction-time limit.

In summary, an experimental measurement of a purely positive \vtwo in pp collisions would rule out the interacting string model, while both hydrodynamics and kinetic theory are consistent with a positive sign. An experimentally confirmed negative \vtwo, on the other hand, would rule out hydrodynamic flow for small systems, confirm a prediction by the interacting string model and could be an indication that interactions take place over sizable distances in kinetic theory.

The sign of \vtwo is thus directly linked to the relevant degrees of freedoms (strings, gluons as well-defined quasi-particles, strong gluon fields, \dots) and their interactions. Our findings, therefore, not only challenge the existing paradigm but also open new avenues for probing the microscopic mechanisms underlying the strong nuclear interaction under extreme conditions.

\subsection*{Acknowledgements}
KZ would like to thank Urs Wiedemann for enlightening discussions. 
Support from the following research grants are gratefully acknowledged. Vetenskapsrådet contracts 2023-04316 (CB), 2020-04869 (LL, GG), 2021-05179 (PC) 2016-05996 (LL, CB, GG, KZ).  Knut and Alice Wallenberg foundation contract number 2017.0036 (CB, PC and LL) and Xunta de Galicia (CIGUS Network of Research Centres), Maria de Maeztu excellence unit grant CEX2023-001318-M, PID2020-119632GB-I00 by MICIU/AEI/10.13039/501100011033, and ERDF/EU (RT). This work has received funding from the European Research Council (ERC) under the European Union's Horizon 2020 research and innovation programme, Grant agreements ERC-2018-StG-803183 (KZ and RT) and ERC-2018-ADG-835105 (RT).

\appendix

\section{Appendix: The Lund string model}
\label{app1}
The Lund string model \cite{Andersson:1979ij,Andersson:1983jt,Andersson:1983ia} is an integral part of the \pythia Monte Carlo event generator \cite{Bierlich:2022pfr}. Throughout this paper we use the implementation of Lund strings in \pythia, which over the recent years has been extended with the possibility of interacting strings in the ``shoving model'' \cite{Bierlich:2017vhg,Bierlich:2020naj,Bierlich:2024odg}, as well as heavy ion initial states in the Angantyr extension \cite{Bierlich:2018xfw}. We will here explain the physics basis of hadronization of a) a single string in isolation, b) a few strings with interactions, such as in the hotspot examples in Figs.~\ref{fig2} and \ref{fig3} and c) in heavy ion collisions, such as in Fig.~\ref{fig4}.

\subsection{Hadronization of a single string}
The main physics picture of the Lund string, is that of a quark--antiquark ($q\bar{q}$) pair, flying apart in one spatial dimension ($x$). Results from Regge theory and lattice QCD supports a linear confining potential between the two, giving rise to the physics description of a massless, relativistic string, with a tension of $\kappa \approx 1$ GeV/fm. This yields a rich kinematical picture \cite{Artru:1979ye} for describing string motion, and is quite general. The Lund string model adds to this, by introducing the concept of gluons as ``kinks'' on the string \cite{Sjostrand:1984ic}, as well as providing a mechanism for breaking the string into smaller pieces, \ie, the hadrons. This can be developed in analogy with the production of $\mathrm{e}^+\mathrm{e}^-$ pairs in a homogenous electric field, the so-called Schwinger mechanism \cite{Schwinger:1962tp}, which in the string case can be understood as the $q\bar{q}$ pair tunneling out of the vacuum. The tunneling probability can be calculated in the WKB approximation \cite{Andersson:1983jt}:
\begin{equation}
\label{eq:schwinger}
    \frac{\mathrm{d}\mathcal{P}}{\mathrm{d}^2p_\perp} \propto \exp(-\pi m^2_q/\kappa)\exp(-\pi p^2_\perp/\kappa),
\end{equation}
with $m_q$ being the mass of the quarks tunneling out of the vacuum when the string breaks, and $p_\perp = \sqrt{p^2_y + p^2_z}$ the quark momentum transverse to the string direction. The Lund symmetric fragmentation function \cite{Andersson:1983jt} provides the kinematics along the string direction as a probability distribution $f(z)$ for taking away a fraction $z$ of the remaining momentum of the string:
\begin{equation}
\label{eq:lundff}
    f(z) \propto \frac{(1-z)^a}{z}\exp(-b(p^2_\perp + M^2)/z),
\end{equation}
where $M$ is the hadron mass and $a$ and $b$ are parameters of the model. This simple picture has shown itself remarkably robust in describing a wealth of experimental data in collisions of $\mathrm{e}^+\mathrm{e}^-$, $\mathrm{e}$p and pp since the 1980s.

\subsection{Hadronization of few-string systems}
When two strings are close to each other, one must take into account their transverse extension. As such, strings will interact with each other, giving rise to flow. The transverse shape of the color-electric field can be calculated in lattice QCD \cite{Cea:2014uja} in the equilibrium limit, and is well approximated by a Gaussian $E(\rho) = N\exp(-\rho^2/2R^2)$, where $\rho$ is the normal radial coordinate, and $R$ the equilibrium radius. Assuming that the energy in the field is a constant fraction $g$ of the total energy $\kappa$, one obtains $N^2 = 2g\kappa/\pi R^2$. In an Abelian approximation\footnote{See Ref.~\cite{Bierlich:2020naj} for a discussion of corrections in the non-Abelian case.}, the repulsive force per unit length becomes:
\begin{equation}
\label{eq:shoving}
    f(d_\perp) = \frac{g \kappa d_\perp}{R^2}\exp(-d^2_\perp/4R^2),
\end{equation}
with $d_\perp$ the distance between two parallel strings. The system evolves in time $t$, and the total applied push becomes $\Delta p_\perp = \int \mathrm{d}t \int \mathrm{d}x f(d_\perp(t))$. Splitting the total push up into a series of smaller pushes, they can be ordered in time using the veto algorithm given an appropriate Lorentz frame \cite{Bierlich:2020naj}.

The above treatment is implemented in the Monte Carlo for the two- and four-string systems studied in the article. The strings are always parallel to each other, and placed with distances to each other according to the description given. The quantity $\Delta p_\perp$ is calculated for each string piece fragmenting into a hadron according to Eqs. (\ref{eq:schwinger})--(\ref{eq:lundff}), and applied in the simulation after the hadron is formed. The $v_2$ is then calculated for all hadrons in the final state.

\subsection{Hadronization with realistic heavy ion final states}
The realistic initial states shown in Fig.~\ref{fig4} are constructed using the Angantyr model \cite{Bierlich:2018xfw}. Nucleons are distributed according to a Woods--Saxon distribution in the cases of $^{197}$Au and $^{12}$C, and using a Hulthén potential \cite{Hulthen:1942} for $^2$d. When nuclei are collided, the amount of nucleons colliding with each other, and their type of interaction, is calculated using a Glauber model \cite{Miller:2007ri}, with nucleon size, $r$, fluctuations modelled by $P(r) = r^{k-1}e^{-r/r_0}/(\Gamma(k)r^k_0$), where $\Gamma$ is the Gamma-function and $r_0$ and $k$ are parameters. Using the Good--Walker formalism \cite{Good:1960ba} parameters can be determined from proton--proton semi-incluive scattering cross sections. Depending on the type of interaction, the sub-collision is then modelled as a different type of minimum bias nucleon--nucleon interaction.

As the sub-collisions will not produce strings which are completely parallel, calculating $d_\perp$ from Eq. (\ref{eq:shoving}) introduces further complexity. In a special Lorentz frame \cite{Bierlich:2020naj}, two string pairs will always lie in parallel planes, and a fractional push, depending on the angle between the strings, can therefore be calculated in this frame.

With realistic initial states, $\Psi$ must be calculated in each event. To allow for the most fine-grained estimate possible, we calculate $\Psi$ from string positions in transverse space at mid-rapidity. Each nucleon--nucleon sub-collision will give rise to a different number of strings, \ie fluctuations in energy density, which will manifest in a varying amount of hadrons being produced from different regions. With positions ($r,\varphi$) of each string known in the simulation, the standard definition is used:
\begin{equation}
    2\Psi_2 = \text{arctan}\left(\frac{\langle r^2\sin(2\varphi) \rangle}{\langle r^2\cos(2\varphi) \rangle} \right) + \pi.
\end{equation}

\section{Appendix: Kinetic theory}
\label{app2}
Kinetic theory is used to predict macroscopic properties and transport phenomena of systems consisting of a large number of particles based on the microscopic interactions between the particles. Below we provide a short explanation of a) a kinetic theory suitable to explain heavy ion collisions and small collision systems, b) a parton cascade implementation of this kinetic theory as well as c) the more realistic initial conditions used in this parton cascade to produce the left hand side of Fig.~\ref{fig4}. 

\subsection{AMY kinetic theory}
In the context of heavy ion collisions the effective kinetic theory of QCD at high temperatures formulated by Arnold, Moore and Yaffe \cite{Arnold:2002zm} has received a lot of attention. At high temperatures QCD is weakly coupled, \textit{i.e.}, $g(T)\ll 1$, and the effective (thermal) masses $m_\text{eff} \sim gT$ are small compared to typical momenta, which are $\mathcal{O}(T)$. The QCD plasma then consists of quarks and gluons as well-defined quasi-particles that propagate nearly freely, since the small angle scattering rate is $\mathcal{O}(g^2T)$. The AMY effective kinetic theory formulates a set of Boltzmann equations for the (anti-)quarks and gluons that describe the change in the phase space densities due to free streaming and interactions,
\begin{equation}
	\left(\partial_t + \frac{\p}{p_0} \cdot \nabla_\mathbf{x} \right) f_s(\x,\p,t) = - C_s^{2\leftrightarrow 2}[f] - C_s^{``1\leftrightarrow 2"}[f].
\end{equation}
The microscopic dynamics is encoded in the collision kernels $C_s$. There are two types of processes, namely elastic scattering ($C_s^{2\leftrightarrow 2}$) and quasi-collinear splitting and merging ($C_s^{``1\leftrightarrow 2"}$).

Generically, a Boltzmann equation based on binary interactions is applicable when the mean free path is long compared to the duration of an interaction. Technically, the finite duration of interactions is neglected so that elastic scattering is assumed to be instantaneous. In the case of the AMY effective kinetic theory this is a good approximation, because it is valid in weakly coupled regimes that support a separation of scales where the momenta of all relevant excitations are large compared to the screening mass, which in turn is large compared to all other mass scales. In particular, near thermal equilibrium this implies that typical momenta are $\mathcal{O}(T)$, the screening mass is $\mathcal{O}(gT)$ and the small angle scattering mean free path is $\mathcal{O}(1/(g^2T))$. The duration of elastic scattering can be estimated as the inverse of the typical momentum transfer, or the invariant mass of the propagator. Taking the latter to be of the order of the screening mass leads to a scattering time $\mathcal{O}(1/(gT))$. Since $g\ll 1$ this is small compared to the mean free path.
We would like to point out that also in out-of-equilibrium situations the differential cross section for small angle scattering is of the form $\mathrm{d}\,\sigma/\mathrm{d}\,t \propto 1/(-t + \mu^2)^2$, where $t$ refers to the Mandelstam variable and $\mu$ is the effective mass. The integrated cross section then goes parametrically as $\sigma \sim 1/\mu^2$, while the interaction time goes like $1/\mu$.  One would then require $1/\mu$ to be smaller than the mean free path.

\subsection{The parton cascade \textsc{ALPACA}}
The phase space density $f_s(\x,\p,t)$ is to be understood as averaged over many configurations or events and the Boltzmann equation is thus an evolution equation for the average. Using parton cascades it is possible to simulate individual events according to the collision kernels of the Boltzmann equation. Taking the average over many events one then recovers the result of evolving the average with the Boltzmann equation. But in addition, parton cascades provide information about fluctuations. In \cite{Kurkela:2022qhn} the parton cascade ALPACA was introduced, which is a parton cascade representation of the AMY effective kinetic theory. In parton cascades a configuration of particles (position and momentum) is sampled from the initial distribution. The time evolution of the system is simulated by allowing the particles to interact according to the differential interaction rates when they get close enough. In the simplest version, the black disk approximation, two particles interact if their distance of closest approach is smaller than $\sqrt{\sigma/\pi}$, where $\sigma$ is the interaction cross section. The particles are thus not at the same point when they interact, which leads to causality violation in relativistic systems. This problem can be avoided by evolving not in time, but in a parameter $\tau$ that is a Lorentz scalar and allows for a frame-independent ordering of interactions \cite{Sudarshan:1981pp,Maruyama:1996rn,Peter:1994yq}. This approach is used in ALPACA, and has been implemented also in other parton cascades \cite{Sorge:1989dy,Borchers:2000wf,Nara:2023vrq}. In addition, a Lorentz-invariant definition of the distance between two particles has to be used.

In a parton cascade particles can interact when their distance is smaller than $\sqrt{\sigma/\pi}$, which can therefore be regarded as the interaction range. Following the argument above the interaction range is $\sqrt{\sigma/\pi} \sim 1/\mu$, which coincides with the duration of  the interaction. In order for the Boltzmann equation to be applicable one thus has to require that $1/\mu$ is small compared to the mean free path.

\subsection{Initial conditions}
The initial phase space distribution used for the results shown in Fig.~\ref{fig4} has the form 
\begin{equation}
    \label{eq:CGClike}
    f = \frac{A}{p_\xi}e^{-\frac{2p_\xi^2}{3}}\left[1 + \epsilon \frac{|\x_\perp|^2}{R_0^2}\cos\left(2\left[\phi_{\mathbf{x}} - \frac{\pi}{2}\right] \right) \right], \quad p_\xi = \frac{\sqrt{p_\perp^2 + \xi^2p_z^2}}{Q(\x_\perp)},
\end{equation}
where $A$ controls the magnitude of the occupancy, $R_0$ corresponds to system size and $\epsilon$ controls the magnitude of the spatial perturbation. The rescaled momentum $p_\xi$ contains a factor $\xi$ which controls the longitudinal momentum asymmetry, and a factor $Q(\x_\perp)$ corresponding to the characteristic energy scale. The latter can be related to the average transverse momentum through $\langle p_\perp^2 \rangle |_{t=t_0, \x_\perp} = Q^2(\x_\perp)$. We chose $Q(\x_\perp)$ as a Gaussian with width $\sqrt{2}R_0$, i.e. $Q(\x_\perp) = Q_0\exp(-|\x_\perp|^2/4R_0^2)$.

As a starting point in \textsc{Alpaca}, we use the initial values $R_0=0.88$ fm, $Q_0 = 0.92 \GeV$ and $\epsilon = 0.33$ derived in \cite{Kurkela:2021ctp} to correspond to a central $pp$ collision system at $\sqrt{s_{\mathrm{NN}}}=5.02 \TeV$ and $\eta/s = 0.2$. However, we do not completely match the energy density to these calculations, but rather let it follow by adjusting the occupancy $A$ to keep the number of initial gluons fixed to $16$. This is done in order to avoid problems with having too few initial particles as we scale our system to smaller system sizes. We focus on a purely gluonic system with only elastic scatterings.


\end{document}